# Vortex Ferroelectric Domains

A. Gruverman,[1] D. Wu,[2] H.-J. Fan,[3] I. Vrejoiu,[4] M. Alexe,[4] R. J. Harrison,[3] and J. F. Scott[3]

[1] Department of Physics and Astronomy, University of Nebraska-Lincoln, Lincoln, NE 68588-0111

[2] Department Materials Science and Engineering, North Carolina State University, Raleigh, NC 27695

[3] Earth Sciences Department, University of Cambridge, Cambridge CB2 3EQ, U. K.

[4] Max Planck Institute of Microstructure Physics, Weinberg 2, D-06120 Halle/Saale, Germany

**Abstract**

We show experimental switching data on microscale capacitors of lead-zirconate-titanate (PZT), which reveal time-resolved domain behavior during switching on a 100-ns scale. For small circular capacitors, an unswitched domain remains in the center while complete switching is observed in square capacitors. The observed effect is attributed to the formation of vortex domain during polarization switching in circular capacitors. This dynamical behavior is modeled using the Landau-Liftshitz-Gilbert equations and found to be in detailed agreement with experiment. This simulation implies rotational motion of polarization in the xy-plane, a Heisenberg-like result supported by the recent model of Naumov and Fu [Phys. Rev. Lett. **98**, 077603 (2007)], although not directly measurable by the present quasi-static measurements.

Over the past 60 years several physicists have considered the possibility that magnetic spins or electric polarization vectors might order not rectilinearly but in circles or toroids. These vortex domain structures (sometimes called "closure domains") in ferroic materials were first predicted by Landau and Lifshitz in 1935 [1] and by Kittel in 1946 [2], who showed that formation of circular domains was likely in ferromagnetic nanodots due to the surface boundary conditions. In 1979, Mermin [3] analyzed their two-dimensional structures in terms of winding numbers, which was extended by Muxworthy et al [4] to three-dimensional "vortex" domains. Such domain patterns are well known in nano-ferromagnets, such as Fe/Ti ilmenite, and are best studied by electron holography [5], but have not been reported in ferroelectrics experimentally, although predicted very recently [6, 7]. The extension to ferroelectrics was made around 1984 by Ginzburg et al [8] and was recently further developed using *ab initio* calculations by Naumov et al [9]. We note parenthetically that toroidal domains can arise for two unrelated physical reasons: (1) finite size effects and boundary conditions (present work); (2) magnetoelectric coupling in multiferroic materials [7, 8, 10, 11, 12]. The present work does not involve magnetism.

Here we report the direct observation of polarization patterns attributed to vortex domains appearing during polarization reversal in ferroelectric capacitors via ultra-fast piezoresponse force microscopy (PFM) [13]. The switching experiments were carried out by using a conducting probing tip, which was in contact with the top electrode deposited on 50-nm thick PZT film. The PZT layer underneath the top electrode was partially switched by applying a short (less than total switching time) voltage pulse. Then the entire capacitor was scanned in the PFM mode to visualize the resulting domain pattern developed in the capacitor. The field was then applied for a longer time interval and the imaging PFM procedure repeated.

Because domain nucleation is nearly 100% inhomogeneous in PZT, occurring at the same defect sites, it was then possible to superimpose the collection of images obtained for different pulse durations to form a slow-motion video of nucleation and domain wall motion during polarization reversal [13]. Domain imaging has been performed through the top electrode by applying an oscillating bias of 0.6 V (peak-to-peak) at 10 kHz. Samples used in this study were 50-nm thick epitaxial (001)-oriented PZT/Pt capacitors with circular 1-µm-diameter and square 2.5-µm-edge Pt top electrodes. Epitaxial $PbZr_{0.2}Ti_{0.8}O_3$ (PZT 20/80) were fabricated by pulsed laser deposition on $SrRuO_3/SrTiO_3$ (001) substrates. Details on film fabrication and basic properties are given elsewhere [14]. Top Pt electrodes have been fabricated by e-beam lithography and lift-off.

The snapshots of instantaneous domain patterns with vertical component of polarization $P_z$ developing in circular and square PZT capacitors during polarization reversal are shown in Fig. 1. It can be seen that in the circular capacitor domains nucleate at the top electrode edge and propagate around the circular circumference forming a doughnut type or "vortex" structure in just a few microseconds (Fig. 1(a)). However, the vortex structure remains for > 1 s, after which the vortex collapses to leave a uniformly polarized ground state. The final collapse follows from Maxwell's equations, particularly

$$-\nabla \times D = dB/dt, \qquad (1)$$

which implies that any vortex is unstable against decay unless $\nabla \times D = 0$. In contrast, the square shape capacitors never exhibit a vortex domain structure (Fig. 1(b)). Instead, switching proceeds via randomly distributed nucleation events with subsequent isotropic lateral domain growth. In general, the observed features of domain kinetics can be summarized as following:

1. The vortex domains occur only in small circular capacitors (<1 μm in diameter) and never in larger square-shaped capacitors.

2. There are always several nucleation sites and all the nucleation occurs near the walls, but none occur at the domain walls.

3. The doughnut type domain structure develops within the time period of several microseconds, lasts for several seconds and then transforms to a single domain state.

Vortex domains are well known in magnetism. Three-dimensional micromagnetic modeling by Muxworthy et al [4] predicted that magnetostatic interactions can generate vortex states in submicrometer size grains which strongly depend on the grain size and anisotropy. The nucleation and evolution of these vortex domain states is unclear due to periodic boundary conditions imposed in the modeling, which obscures the inhomogeneous surface nucleation process in real materials. Note that in ferroelectrics, polarizations do not generally rotate continuously in space, so that the structures shown in [4] with magnetization vectors curling around a central core are only qualitative in the description of ferroelectrics. The ferroelectric case can be approximated in magnetic models by letting the anisotropy/exchange ratio become very large. The metastable nature of the vortex domain state was not emphasized by Muxworthy et al [4] but is probably another reason why vortex domains have not been reported before in ferroelectrics: they are only metastable, lasting a few seconds. Note however that this is ten million times longer than the switching time required to produce the state.

The simulation of ferroelectric vortex domain nucleation and switching has been performed using the LLG (Landau-Liftshitz-Gilbert equation) User Manual [15], which in turn is based upon the micromagnetic equations of motion for a Heisenberg magnet as developed by Scheinfein et al [16]. The time evolution of a magnetization configuration described by the Landau-Liftshitz-Gilbert equation has been solved with the pseudo-spin configuration relaxed iteratively. For the application to ferroelectric PZT we adjusted the pseudo-spin parameters such that the anisotropy is larger than in ferromagnets and the exchange energy smaller. The LLG simulation allowed us to model all the details of the experimental data on domain switching (Fig. 2):

1. The vortex domains occur only in small disks and never in square-shaped capacitors.

2. There are always several nucleation sites.

3. All the nucleation occurs near (say 5 lattice constants in the real crystal) from the walls, but none occur at a wall.

4. The switching occurs in approximately 100 time-steps and the doughnut-type center remains for approximately $5 \times 10^6$ time steps and then it vanishes. Our time steps thus each represent ~200 ns in the experiments.

One obvious limitation of the present experiments and models should be mentioned here. In the LLG simulation we watch the actual dynamical switching process in detail, while the domain walls are moving. This process involves not only a simple Ising-like up and down flipping of polarization but also a development of in-plane component of polarization, which arises prior to the complete polarization reversal. However, it is difficult to verify this via direct PFM experiment for at least two reasons. First, the actual domain switching measurements using PFM are quasi-static. The PZT capacitors are partially switched and allowed to remain in this state for a time required for PFM domain image acquisition, which is much longer than the switching time. On the other hand, the vortex-type domains represent non-steady polarization state with the in-plane components appearing only during dynamic switching. Second, during PFM measurements through the top electrodes, the lateral PFM signal is inevitably affected by the contribution of $d_{31}$ and $d_{32}$ constants of the piezoelectric tensor of PZT, which makes unambiguous detection of the in-plane polarization difficult [17].

Thus, the line of inference is that the numerical simulation very closely mimics the $P_z$ polarization patterns as a function of time, and that the dynamics in the numerical simulation closely resembles the in-plane polarization model of Naumov and Fu [7]. Furthermore, recent results by Junquera strongly support our modeling [18].

A cylindrical waveguide and Bessel function analogy may be useful to explain both the experimentally observed domain patterns and the switching times. Let us consider the circular PZT capacitors to be cylindrical waveguides. In this case, the lower TEM modes are well known, and they look like the patterns we see. So we might treat the domain wall nucleation and growth in PZT capacitors as occurring at the high-field regions of a circular waveguide of radius $B$. The doughnut-shape domain can be described by an internal radius $a$ and an external radius $b$, with $b<B$. One finds that the impedance $Z$ of the transmission "line" is the ratio of voltage between the inside and outside to the axial current flowing normal to the disk. In our case the axial current $J$ is not ohmic but instead a displacement current, so borrowing the waveguide results [19] we can write

$$dP/dt = J = (2n/c)V \ (\mu/\varepsilon)^{1/2} \ln (b/a) \qquad (2)$$

where $\varepsilon$ is the dc current $\varepsilon(0)$ and not $\varepsilon(\infty) = n^2$. Although this approach is only qualitative and does not yield quantitative agreement with the measured values of *dP/dt*, the idea of replacing real current *J(t)* with displacement current *i(t)* = *dP/dt* in ferroelectric nano-rods during switching may merit further consideration.

A simpler but analogous description for the spatial patterns observed in Figs. 1 and 2 is that they resemble Bessel functions which are used for description of the propagation of electric currents in cylindrical conductors [20, 21, 22].

In summary, we report for the first time the polarization patterns during switching in micron-diameter ferroelectric disks on a 100 ns time scale. The observed domain pattern with an unswitched circular domain at its center is attributed to a vortex domain which develops during dynamic switching. This vortex domain does not occur for square ferroelectrics of the same size. Numerical switching simulations using the Landau-Lifshitz-Gilbert equations agree very closely with experiment for both disks and squares. The simulations also predict a significant in-plane polarization during switching, which agrees with the model of Naumov and Fu [7] but is beyond present experimental detection.

**Figure captions**

Fig. 1. PFM images of instantaneous domain configurations with out-of-plane polarization component $P_z$ developing at different stages of polarization reversal in (a) 1-µm-diameter circular capacitor, (b) 2.5-µm edge square capacitor. Applied electric field is 700 kV/cm. The total switching time for the domain in the center of the disk is about 1 s.

Fig. 2. Simulation of PZT switching: (a) 1-µm-diameter circular capacitor; (b) 2.5-µm edge square capacitor. The top trace shows the time evolution of out-of-plane polarization component $P_z$; Lower traces illustrate time evolution of in-plane polarization components $P_x$ and $P_y$, which cannot be measured with the present PFM technique.

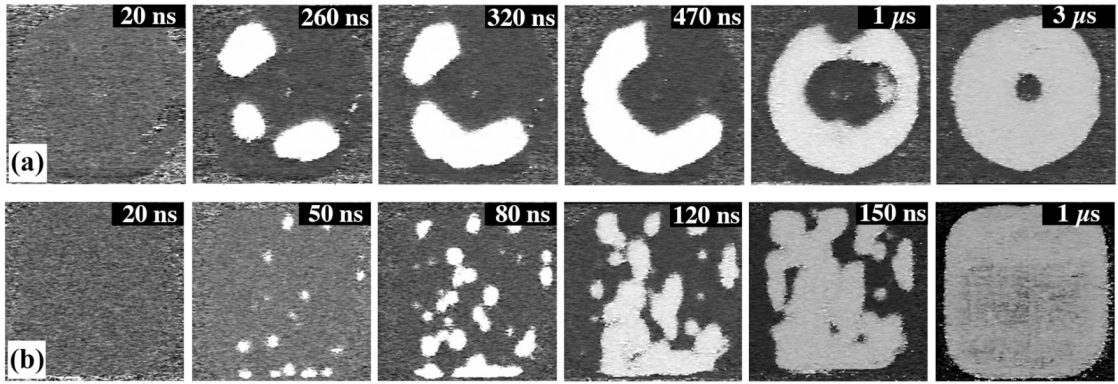

Fig. 1.

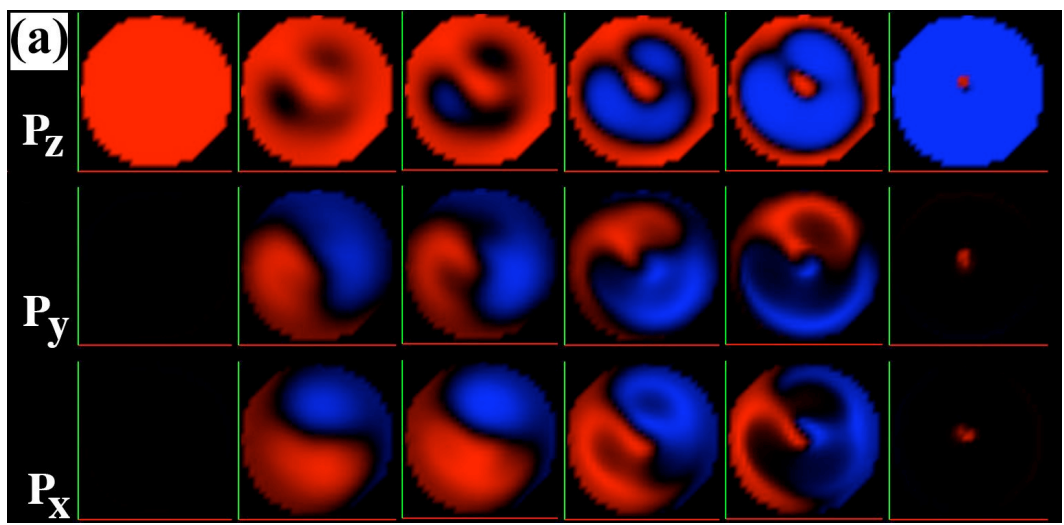

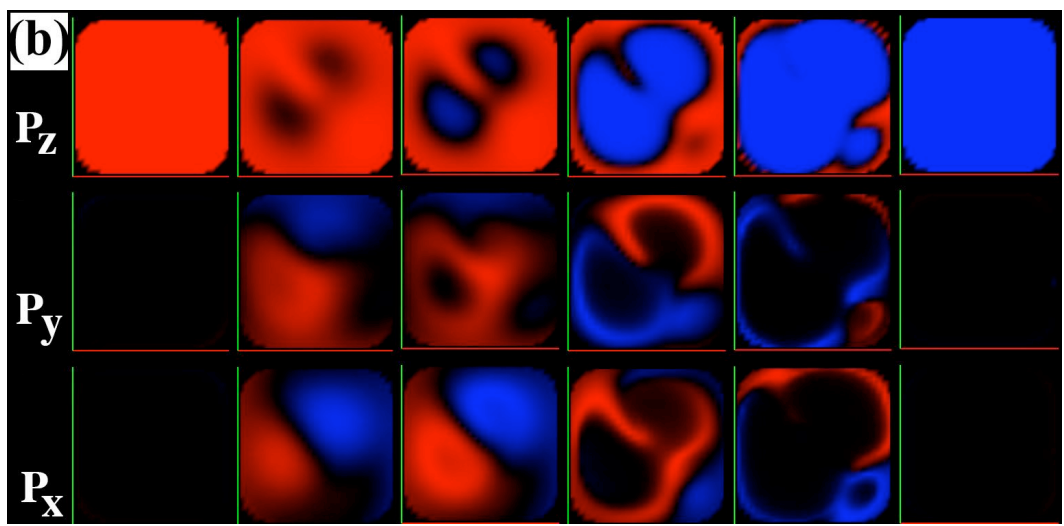

Fig. 2

[20] I. S. Sokolnikoff and R. M. Redheffer, *Mathematics of Physics and Modern Engineering* (McGraw-Hill, New York, 1958), p.159.

[21] U. A. Bakshi and A. V. Bakshi, *Transmission Lines and Waveguides* (Technical Publications, Pune, 2006), Chap. 5.

[22] N. Cronin, *Microwave and Optical Waveguides* (Adam Hilger, Bristol, 1995); Chap. 6.